\documentclass{article}
\usepackage{epsfig}
\begin{document}
\begin{large}
\baselineskip 0.26in

\titlepage
\vspace*{0.4in}

\begin{center}
\begin{LARGE}
{\bf An Equation-Free Approach to Nonlinear Control: Coarse
Feedback Linearization With Pole-Placement}
\end{LARGE}
\vspace{0.20in}
\end{center}

\begin{center}

{\bf Constantinos I. Siettos \footnote{\it Currently: School of
Applied Mathematics and Physics, National Technical University of
Athens, Zografou Campus, Athens, GR 157 80, Greece}}\\
\vspace{0.1in} \it Department of Chemical Engineering, Princeton
University,
Princeton, NJ 08544\\
Electronic mail address: ksiet@mail.ntua.gr\\

\vspace{0.4in}

{\bf Ioannis G. Kevrekidis\footnote{Author to whom correspondence should be directed}}\\
\vspace{0.1in} Department of Chemical Engineering, Department of
Mathematics and Program in Computational and Applied Mathematics,
Princeton University, Princeton, NJ 08544\\
Electronic mail address: yannis@princeton.edu \\

\vspace{0.4in}

{\bf Nikolaos Kazantzis} \\
\vspace{0.1in} Department of Chemical Engineering, Worcester
Polytechnic Institute, Worcester,  MA  01609-2280,
Electronic mail address: nikolas@wpi.edu\\

\date{}
\vspace{0.4in}

\begin{abstract}
We present an application of equation-free computation to the
coarse-grained feedback linearization problem of nonlinear systems
described by microscopic/stochastic simulators.
Feedback linearization with pole placement requires the solution
of a functional equation involving the macroscopic
(coarse-grained) system model.
In the absence of such a closed-form model, short, appropriately
initialized bursts of microscopic simulation are designed and
performed, and their results used to estimate {\it on demand} the
quantities required for the numerical solution of the (explicitly
unavailable) functional equation.
Our illustrative example is a kinetic Monte Carlo realization of a
simplified heterogeneous catalytic reaction scheme.

\end{abstract}
\end{center}

\newpage

\section{Introduction}

A fundamental prerequisite for the design of control systems is
the availability of reasonably accurate closed form dynamical
models.
Typically, such models arise in the form of evolution equations
(ordinary differential, differential algebraic, partial
differential, possibly integrodifferential equations).
Such equations are typically derived from conservation laws (e.g.
mass, momentum and energy balances) closed through constitutive
equations (e.g. Newtonian stresses in fluid flow, or mass-action
kinetics expressions for chemical reactions); system
identification may also play a role in obtaining and/or closing
such macroscopic models.
Many real-world  problems of current engineering interest are
characterized -due to their stochastic/microscopic nature, and
nonlinear complexity- by the lack of such good explicit,
coarse-grained macroscopic evolution equations.
Instead, the underlying physics description may be available at a
much {\it finer}, more detailed level: the evolution rules may be
given in the form of molecular dynamics, kinetic Monte Carlo,
Markov-chain or hybrid schemes.
When this is the case, conventional continuum algorithms cannot be
used directly for systems level analysis and controller design.
Bridging systematically the enormous gap between microscopic space
and time scales of a complex physical/material system description
and the macroscopic ones at which we want to design and control
its behavior is a grand challenge for modeling and computation.
Over the past few years we have demonstrated that an equation-free
approach (based on coarse timesteppers) [Theodoropoulos et al.,
2002; Makeev et al.,2002; Kevrekidis et al., 2003; Siettos et al.,
2003b; Kevrekidis et al., 2004], can establish a link between
traditional continuum numerical analysis and microscopic/
stochastic simulation.
This is a mathematics-assisted computational methodology, inspired
from continuum numerical analysis, system identification and large
scale iterative linear algebra, which enables microscopic-level
codes to perform system-level analysis directly, without the need
to pass through an intermediate, coarse-grained,
macroscopic-level, ``conventional" description of the system
dynamics.
The backbone of the method is the {\it on-demand} identification
of the quantities required for continuum numerics (coarse
residuals, the action of coarse slow Jacobians, eigenvalues,
Hessians, etc).
These are obtained by repeated, appropriately initialized calls to
an existing {\it fine scale} time-stepping routine, which is
treated as an input-output black box.
The key assumption is that deterministic, macroscopic, coarse
models exist and close for the expected behavior of a few
macroscopic system observables, yet they are unavailable in closed
form.
These observables (coarse-grained variables) are typically a few
low moments of microscopically evolving distributions (e.g.
surface coverages, the zeroth moments of species distributions on
a lattice model of a surface reaction).

The present work aims at developing a systematic approach to the
feedback regulator synthesis problem, where both the closed-loop
dynamics linearization and pole-placement objectives are
simultaneously attained by using the equation-free timestepper
methodology. The  feedback linearization and the pole-placement
objectives for the unavailable coarse-grained dynamics are met in
a {\it single-step}, circumventing the lack of an explicit dynamic
process model.
The proposed approach is illustrated through the use of a coarse
time-stepper based on a kinetic Monte Carlo realization of a
simplified surface reaction scheme for the dynamics of $NO$
oxidation by $H_2$ on $Pt$ and $Rh$ surfaces. The present paper is
organized as follows: In section 2 we briefly discuss the
traditional nonlinear control methodologies that rely on the
notion of feedback linearization along with the associated
restrictions encountered at the implementation stage. In section 3
we succinctly review a recently proposed approach that allows the
attainment of both the feedback linearization and pole placement
objectives in a single step, effectively overcoming the
restrictive conditions associated with the classical exact
feedback linearization approach. In section 4 the interplay of the
proposed nonlinear control procedure with coarse timesteppers is
outlined, and the natural integration of the respective frameworks
illustrated. Section 5 presents the simulation results using the
proposed methodology on an illustrative kinetic Monte Carlo model,
followed by some concluding remarks in section 6.

\section {Fundamentals of feedback linearization of nonlinear systems}

In order to meet a set of performance specifications or design
objectives, process control introduces feedback to appropriately
modify the dynamics of a system.
Placing the closed-loop poles at desirable locations in the
complex plane, and thus shaping the closed loop system dynamics
and time constants, is a popular controller synthesis method for
linear systems, in part, due to its intuitive appeal [Chen, 1984].
Typically one requires fast decay of the closed loop variables to
their nominal steady state values; yet the design should not lead
to high feedback gains due to possible saturation problems.
Fine-tuning of the closed-loop  eigenvalues is performed in
practice through a combination of optimization techniques,
heuristic rules and trial-and-error approaches [Chen, 1984].
Traditional pole-placement state feedback control laws for
nonlinear systems are based on local linearization around a
reference steady state, and the subsequent use of linear design
methods.
The results are, of course, only locally valid, and may lead to
unacceptable performance, even in the presence of only mild
nonlinearities.
Nonlinear feedback control laws thus need to be derived, capable
of directly coping with the system nonlinearities.
A pole-placing feedback regulator should be capable of bringing
the system/process state back to the design steady state in a fast
and smooth manner in the presence of disturbances; if the design
steady state is unstable, the primary control objective is its
stabilization.
In the pertinent body of literature two main model-based
pole-placing controller synthesis methods emerge, both based on
geometric control theory.
The first one is exact input/output (I/O) feedback linearization,
where the introduction of nonlinear state feedback induces linear
I/O behavior of the system of interest, forcing the system's
output to follow a prespecified linear and stable trajectory.
This approach directly generalizes the linear result of placing
the closed-loop poles at the system's zeros and at a set of
prespecified values, and is restricted within the class of
minimum-phase systems [Isidori, 1999].
Regulation and/or stabilization of a system/process, however, is
understood in terms of forcing the system's state to return to the
design steady state (if driven away from it in the presence of
disturbances). Furthermore,  process output tracking problems for
step changes in the output set-point values, can be easily
reformulated as regulation problems relative to the equilibrium
point that corresponds to the final set-point value.
The second  approach is geometric exact feedback linearization,
traditionally implemented in a two-step design procedure [Isidori,
1999]: A simultaneous implementation of a  nonlinear coordinate
transformation and a state feedback control law in the first step
transforms the original nonlinear  system to a linear and
controllable one.
Well-established linear pole-placement techniques for the
transformed linear system can be used in the second step.
However, the aforementioned classical geometric exact feedback
linearization approach relies on a set of very restrictive
conditions, that can hardly be met by any physical system.

In this work a systematic approach to feedback regulator synthesis
is proposed for the coarse-grained dynamic behavior of systems
described by atomistic/stochastic (``fine scale") simulators.
The closed-loop dynamics linearization and the pole-placement
objectives are simultaneously attained using the equation-free
timestepper-based methodology.
Note that our primary control objective is to assign the
closed-loop eigenvalues rather than shaping the entire I/O
behavior of the system under consideration. Furthermore, applying
the methodology introduced in [Kazantzis, 2001], we investigate
the possibility of circumventing the set of restrictive conditions
associated with the two-step classical exact feedback
linearization approach, by meeting the feedback linearization and
the pole-placement objectives in a {\bf single-step}, and without
being limited by the availability of an explicit dynamic process
model.

\section{Mathematical Preliminaries - Problem Formulation}
In the context of the present study, the system dynamics are
described by a nonlinear discrete-time macroscopic (``coarse")
model of the form:
\begin{equation}
x(k+1)=\Phi(x(k),u(k)).
\end{equation}
Here $k \in N^{+}=\{0,1,...\}$ is the discrete-time index, $x(k)
\in R^{n}$ is the vector of (coarse) state variables, $u(k) \in R$
is the manipulated input variable and $\Phi(x,u)$ represents a
vector function defined on $R^{n} \times R$.
In our case this function is not known, and will be identified on
the fly with the aid of the fine scale simulator.
Without loss of generality, it is assumed that the origin
$x^{0}=0$ is an equilibrium point (coarse steady state) of (1),
that corresponds to $u^{0}=0$: $\Phi(0,0)=0$.
If a non-zero coarse steady-state $(x^{0},u^{0}) \neq (0,0)$ is
located, then a simple transformation:
$\hat{x}=x-x^{0},\hat{u}=u-u^{0}$ will map it to the origin in the
new coordinate system.
Let $F$ be the Jacobian matrix of $\Phi(x,u)$ evaluated at $x=0$:
$\displaystyle{F=\frac{\partial \Phi}{\partial x}(0,0)}$, and $G$
the vector: $\displaystyle{G=\frac{\partial \Phi}{\partial
u}(0,0)}$ which is assumed to be non-zero.
The following assumption is also made:

{\bf Assumption I:} The $n \times n$ matrix:
\begin{equation}
{\cal{C}}=\left[\matrix{G|FG|...|F^{n-1}G}\right]
\end{equation}
has rank $n$.
This implies that the coarse linearization of (1) around the
origin $x=0$ is controllable [Isidori, 1999].

It is appropriate, at this point, to briefly review and outline
basic features of the classical exact feedback linearization
approach in the discrete-time domain.
In the first step, and under a set of rather restrictive
conditions [Aranda-Bricaire et al., 1996; Califano et al., 1999;
Grizzle, 1986; Jacubczyck, 1987; Lee et al., 1987; Lin and Brynes,
1995; Nam, 1989], a nonlinear coordinate transformation: $z=T(x)$
is sought along with a state feedback control law: $u=\Psi(x,v)$
(with $v$ being an external reference input), such that the
original system (1) is transformed to the following linear one:
\begin{equation}
z(k+1)=Az(k)+bv(k)
\end{equation}
where $(A,b)$ is a Brunowsky controllable pair of  matrices [Chen,
1984].
 In the second step, standard linear pole-placing feedback
techniques are used to arbitrarily assign the poles (equivalently
the time-constants) of the closed-loop system.
In particular, a constant-gain vector $K$ is calculated, such that
the state feedback law: $v=-Kz$ induces the desirable closed-loop
dynamics:
\begin{equation}
z(k+1)=\bar{A}z(k)=(A-bK)z(k)
\end{equation}
with $\bar{A}=A-bK$ being the closed-loop system's characteristic
matrix with prescribed eigenvalues.

At this point it would be appropriate to review an alternative
{\it single-step} design method for linear systems:
\begin{equation}
x(k+1)=Ax(k)+bu(k),
\end{equation}
where $A,b$ are constant matrices with appropriate dimensions,
that was first introduced by D. Luenberger (1963). This
alternative approach serves as the methodological basis for the
development of a nonlinear analogue introduced in [Kazantzis,
2001] and briefly outlined in the next section. According to the
ideas reported in [Luenberger, 1963] a single-step simultaneous
implementation of a linear coordinate transformation: $z=Tx$
coupled with a linear state feedback control law: $u=-Kz$ is
sought, that induce the following closed-loop dynamics:
\begin{equation}
z(k+1)=\bar{A}z(k)
\end{equation}
$\bar{A}$ is the closed-loop system's characteristic matrix that
carries the prescribed set of eigenvalues due to the control law
applied.
This requirement can be mathematically translated into a quadratic
matrix equation that the unknown transformation matrix $T$ should
satisfy:
\begin{equation}
TA-\bar{A}T=TbKT
\end{equation}
If $T$ is non-singular (invertible), one can easily show that the
inverse transformation matrix $W=T^{-1}$ satisfies the following
linear matrix equation:
\begin{equation}
AW-W\bar{A}=bK.
\end{equation}
It is known from linear algebra that, if matrices $A$ and
$\bar{A}$ have disjoint eigenspectra, the above matrix equation
(8) admits a unique solution $W$ [Chen, 1984; Gantmacher, 1960].
Furthermore, invertibility of the solution can be ensured iff the
pair of matrices $(A,b)$ is controllable and the pair
$(K,\bar{A})$ is an observable one [Chen, 1984].
As shown in [Luenberger, 1963], if $T$ is the unique invertible
solution to the matrix equation (7), then the linear state
feedback control law expressed in the original variables $x$
\begin{equation}
u(k)=-KTx(k)
\end{equation}
induces the closed-loop dynamics:
\begin{equation}
x(k+1)=T^{-1}\bar{A}Tx(k)
\end{equation}
Since matrices $T^{-1}\bar{A}T$ and $\bar{A}$ are similar, it can
be easily inferred that the closed-loop system has the desirable
set of poles assigned by the control law (9).

\subsection{Single-Step Feedback Linearization With Pole-Placement}

Motivated by D. Luenberger's linear approach [Luenberger, 1963],
let us now succinctly review the ideas presented in [Kazantzis,
2001] on its nonlinear generalization. One seeks to simultaneously
implement a nonlinear coordinate transformation, $z=S(x)$ coupled
with a nonlinear state feedback control law, $u=-cz=-cS(x)$, where
$c$ is an arbitrary constant row vector (a design parameter of the
proposed method) that induce linear closed-loop $z$-dynamics:
\begin{equation}
z(k+1)=Az(k).
\end{equation}
The poles of the closed-loop dynamics (11) are realized by the
eigenvalues of the arbitrarily prescribed matrix $A$: the
characteristic matrix of the linear closed-loop dynamics (11).
Therefore, the eigenspectrum of $A$ should be judiciously selected
to favorably shape the dynamic characteristics of the controlled
system's response.
In the nonlinear case, these design requirements are embodied into
the following system of nonlinear functional equations (NFEs) that
need to be satisfied by the unknown transformation map $S(x)$:
\begin{eqnarray}
S(\Phi(x,-cS(x)))&=&AS(x) \nonumber \\
S(0)&=&0.
\end{eqnarray}
The accompanying initial condition $S(0)=0$ merely reflects the
fact that equilibrium properties must be preserved under the
proposed coordinate transformation.

For the study of the mathematical properties of the solution of
the NFEs (12) and within the class of real analytic systems, a
number of assumptions are essential as shown in [Kazantzis, 2001]:

{\bf Assumption II:} All the eigenvalues $k_{i}, (i=1,...,n)$ of
matrix $A$ should lie  inside the unit disc on the complex plane
(stability requirement imposed on the closed-loop dynamics (11)).

{\bf Assumption III:} The eigenspectra $\sigma(A),\sigma(F)$ of
matrices $A$ and $F$ respectively should be disjoint: $\sigma(A)
\cap \sigma(F) = \emptyset$.

{\bf Assumption IV:} The eigenvalues $k_{i}$ of $A$ should not be
related to the eigenvalues $\lambda_{j},  (j=1,...,n) $ of F
through any equations of the type:
\begin{equation}
\prod_{i=1}^{n}k_{i}^{m_{i}}=\lambda_{j}
\end{equation}
$(j=1,...,n)$, where all the $m_{i}$'s are non-negative integers
that satisfy the condition:
\begin{equation}
\sum_{i=1}^{n}m_{i}>0.
\end{equation}

{\bf Assumption V:} The pair of matrices $(c,A)$ is chosen such
that the following matrix $O$:
\begin{equation}
O=\left[\matrix{c\cr
                 cA\cr
                 .\cr
                  .\cr
                 cA^{n-1}}\right]
\end{equation}
has rank $n$: rank$(O)=n$ (Observability condition imposed on
$(c,A)$).

{\bf Lemma:} [Kazantzis, 2001] {\it For a real analytic system
(1), let the above assumptions I-V hold true. Then, the  system of
NFEs (12) admits a unique locally analytic and invertible solution
$z=S(x)$ in a neighborhood of the origin $x=0$}.

We include here a number of remarks discussing the conditions and
implications of this Lemma; a more detailed discussion can be
found in [Kazantzis, 2001].

{\bf Remark 1}: The ``non-resonance" conditions (13) and (14) are
required for the existence of a unique formal power-series
solution to the system of NFEs (12).
The assumption for the eigenspectrum of matrix $A$ to lie inside
the unit disc plays a key role in the uniform convergence of this
formal power-series solution in the neighborhood of the origin
$x=0$ with a non-zero radius of convergence, and thus for the
solution's analyticity.
Finally, Assumptions I and V are necessary and sufficient
conditions for local invertibility of the solution.

{\bf Remark 2}: It is useful to consider the linear case:
$\Phi(x,u)=Fx+Gu$ where $F,G$ are a constant matrix and vector of
appropriate dimensions respectively.
In this case, the unique solution of the system of NFEs (12) is
$w={\cal{S}}x$, where $\cal{S}$ is the solution to the quadratic
matrix equation:
\begin{equation}
{\cal{S}}F-A{\cal{S}}={\cal{S}}Gc{\cal{S}}.
\end{equation}
which coincides with (7) in D. Luenberger's analysis. Please
notice, that under the assumptions stated the above matrix
equation (16) admits a unique and invertible solution ${\cal{S}}$
[Chen, 1984].

Let us now consider: $z=S(x)$ to be the solution to the associated
system of NFE's (12) defined in a neighborhood of $x=0$.
It has been shown in Kazantzis (2001) that the simultaneous
implementation of the nonlinear coordinate transformation:
$z=S(x)$ and the nonlinear state feedback control law:
\begin{equation}
u(k)=-cS(x(k))
\end{equation}
results in linear closed-loop $z$-dynamics:
\begin{equation}
z(k+1)=Az(k)
\end{equation}
whose poles are realized by the eigenvalues of matrix $A$.
Indeed, one can easily show that the  closed-loop system dynamics
expressed in the $z$-coordinates satisfy:
\begin{eqnarray}
z(k+1)&=&S(x(k+1))=S(\Phi(x(k),-cS(x(k))
\nonumber \\
&=&AS(x(k))=Az(k).
\end{eqnarray}

{\bf Remark 3}: Note that in the linear case, one calculates a
feedback control law:
\begin{equation}
u(k)=-c{\cal{S}}x(k)
\end{equation}
where $\cal{S}$ is the solution to (16), that induces the
following closed-loop dynamics:
\begin{equation}
x(k+1)=(F-Gc{\cal{S}})x(k)
\end{equation}
Using equation (16), the closed-loop dynamics (21) can be
rewritten as follows:
\begin{equation}
x(k+1)=({\cal{S}}^{-1}A{\cal{S}})x(k)=\tilde{A}x(k).
\end{equation}
Notice that $A,\tilde{A}={\cal{S}}^{-1}A{\cal{S}}$ are similar
matrices, and therefore, the closed-loop system (22) has the
desirable poles.
One can consider this approach as the natural extension of D.
Luenberger's linear result for pole-placement (10) to nonlinear
systems.

{\bf Remark 4}: The graph of the mapping $z=S(x)$ is rendered
invariant for the composite system (1) and (11) under the state
feedback control law: $u(k)=-cS(x(k))$ [Carr, 1981].
Furthermore, the system of NFEs (12) represent the associated
invariance functional equations for the composite system (1)-(11)
[Guckenheimer and Holmes, 1983], and the restriction of the
composite system dynamics under the above feedback law on the
invariant manifold/solution of (12) coincides with the linear
closed-loop dynamics (11).

{\bf Remark 5}: The primary idea of the proposed single-step
design approach is to avoid the intermediate step of transforming
the original system into a linear controllable one with an
external reference input, which allowed us to circumvent the
restrictive conditions associated with the classical exact
feedback linearization method [Aranda-Bricaire, 1996; Califano,
1999; Grizzle, 1986; Jakubczyck, 1987; Lee, 1987; Lin and Byrnes,
1995; Nam, 1989].
It should be pointed out, that the design method described does
not involve an external reference input, however, and therefore
other control objectives such as trajectory tracking or model
matching can not be met [Isidori, 1999].

In the present study the NFEs (12) will be solved using the
equation-free computational framework.
However, for completeness and comparative accuracy, one needs to
employ a alternative solution scheme/method for the system of
NFE's (12).
This method involves expanding $\Phi(x,u)$ as well as the unknown
solution $S(x)$ in a Taylor series and equating the Taylor
coefficients of the same order of both sides of the NFE's (12).
This procedure leads to linear recursion formulas, through which
one can calculate the $N$-th order Taylor coefficients of $S(x)$,
given the Taylor coefficients of $S(x)$ up to the order $N-1$.
As shown in [Kazantzis, 2001], in the derivation of the recursion
formulas, it is convenient to use the following tensorial
notation:

a) The entries of a constant matrix $A$ are represented as
$a_{i}^{j}$, where the subscript $i$ refers to the corresponding
row and the superscript $j$ to the corresponding column of the
matrix.

b) The partial derivatives of the $\mu$-th component
$\Phi_{\mu}(x,u)$ of the vector function $\Phi(x,u)$ with respect
to the state variables $x$ evaluated at $(x,u)=(0,0)$ are denoted
as follows:
\begin{eqnarray}
f_{\mu}^{i}&=&\frac{\partial \Phi_{\mu}}{\partial x_{i}}(0,0) \nonumber \\
f_{\mu}^{ij}&=&\frac{\partial^{2} \Phi_{\mu}}{\partial x_{i}
\partial x_{j}}(0,0)
\nonumber \\
f_{\mu}^{ijk}&=&\frac{\partial^{3} \Phi_{\mu}}{\partial x_{i}
\partial x_{j}
\partial x_{k}}(0,0)
\end{eqnarray}
etc., where $i,j,k,..$=$1,...,n$

c) The partial derivatives of the $\mu$-th component
$\Phi_{\mu}(x,u)$ of the vector function $\Phi(x,u)$ with respect
to the input variable $u$ evaluated at $(x,u)=(0,0)$ are denoted
as follows:
\begin{equation}
g_{\mu}^{i}=\frac{\partial^{i} \Phi_{\mu}}{\partial u^{i}}(0,0)
\end{equation}
etc.

d) The standard summation convention where repeated upper and
lower tensorial indices are summed up.

Under the above notation the $l$-th component $S_{l}(x)$ of the
unknown solution $S(x)$ can be expanded in a multivariate Taylor
series as follows:
\begin{eqnarray}
S_{l}(x)&=&\frac{1}{1 !}S_{l}^{i_{1}}x_{i_{1}}+\frac{1}{2
!}S_{l}^{i_{1}i_{2}}x_{i_{1}}x_{i_{2}}+...+ \nonumber \\
&+&\frac{1}{N
!}S_{l}^{i_{1}i_{2}...i_{N}}x_{i_{1}}x_{i_{2}}...x_{i_{N}}+...
\end{eqnarray}
Similarly one expands the components of the vector function
$\Phi(x,u)$ in multivariate Taylor series. Substituting the Taylor
expansions of $S(x)$ and $\Phi(x,u)$ into (12) and matching the
Taylor coefficients of the same order, the following relation for
the $N$-th order terms may be obtained [Kazantzis, 2001]:
\begin{equation}
\sum_{L=1}^{N}\sum_{0 \leq m_{1} \leq m_{2} \leq ...\leq m_{L}
\atop m_{1}+m_{2}+...+m_{L}=N}S_{l}^{
j_{1}...j_{L}}(f^{m_{1}}_{j_{1}}...f^{m_{L}}_{j_{L}}-\pi^{m_{1}}_{j_{1}}...\pi^{m_{L}}_{j_{L}})=a_{l}^{\mu}S_{\mu}^{i_{1}...i_{N}}
\end{equation}
where:
\begin{equation}
\pi_{j_{l}}^{m_{L}}=\sum_{P=1}^{L} \sum_{0 \leq n_{1} \leq n_{2}
\leq ...\leq n_{P} \atop n_{1}+n_{2}+...+n_{P}=m_{L}}
g_{j_{l}}^{n_{1}} c^{k}S_{k}^{n_{2}...n_{P}}
\end{equation}
$i_{1},...,i_{N}=1,...,n$ and $l=1,...,n$. Notice that the second
summation symbol in (26) (and similarly in (27)) suggests summing
up the relevant quantities over the
$\displaystyle{\frac{N!}{m_{1}!...m_{L}!}}$ possible combinations
to assign the $N$ indices $(i_{1},...,i_{N})$ as upper indices to
the $L$ positions $\{f_{j_{1}},...f_{j_{L}}\}$ (and
$\{\pi_{j_{1}},...\pi_{j_{L}}\}$), with $m_{1}$ of them being put
in the first position, $m_{2}$ of them in the second position ,
etc. ($\displaystyle{\sum_{i=1}^{L}}m_{i}=N)$.
Moreover, notice that equations (26,27) represent a set of linear
algebraic equations in the unknown coefficients
$S_{\mu}^{i_{1},...,i_{N}}$ for $N \geq 2$. For $N=1$, equations
(26,27) yield the quadratic matrix equation (16) (or (7) in D.
Luenberger's approach).
It should be pointed out, that the above series solution method
for the NFE's (12) is amenable to a computer-based implementation
and can be readily carried out in an automatic fashion with the
aid of a symbolic software package such as MAPLE.

\section{An Equation-Free Approach to the Feedback Linearization Problem}

As shown in the previous section, the system of NFEs (12) admits a
unique analytic solution.
However, such an analytic transformation is difficult to derive in
the general case, and a numerical solution scheme becomes
necessary.
We will now assume that the model equations are not explicitly
available, but we do have a ``black box" subroutine that, given
the state of the system $x_{0} \in R^{n}, u_0 \in R $ at time
$t_{k} = kT$ reports the result of the system after a time horizon
T (i.e., will report $x(t_{k+1} = (k+1)T) \equiv
\Phi_T(x_0,u_0)$).
This subroutine could be a ``legacy" dynamic simulator;
alternatively, it can be a ``coarse timestepper" involving the
lift, run and restrict steps discussed briefly below and in more
detail in [Makeev et al., 2002; Gear et al., 2002; Kevrekidis et
al. 2003, Siettos et al., 2003b].
The coarse timestepper, which we use in the equation-free
framework for coarse-grained controller design (for linear
quadratic control, pole placement and feedback linearization)
[Siettos et al., 2003a, Siettos, et al., 2004a, Armaou et al.,
2004a, 2004b, Siettos et al., 2004b] consists of the following
elements (Figure 1):

\begin{itemize}

\item  a lifting operator $\mu$, transforming a macroscopic
initial condition (typically zeroth- or first-order moments of the
microscopically evolving distributions) to one (or more)
consistent microscopic realizations;

\item  evolution of the microscopic realizations using the
microscopic simulator for an appropriately chosen (relatively
short) macroscopic time $T$, ({\it the reporting horizon}).

\item a restriction operator $M$, transforming the resulting
microscopic distributions back to the macroscopic description
(obtaining their macroscopic observables).
Lifting from microscopic to the macroscopic and then restricting
again should have theoretically no effect (modulo roundoff), that
is, $\mu M = I$.

\end{itemize}

This coarse timestepper, appropriately initialized and executed
can serve in the ``on demand" estimation of model
right-hand-sides, of the action of ``coarse slow" Jacobians as
well as derivatives with respect to parameters, in the computation
of coarse fixed points and their leading eigenvalues -- in short,
of exactly the quantities that a linear or nonlinear controller
design algorithm would need evaluated through a macroscopic model
(had such a model been available) to perform its task.

For our problem, we use the coarse timestepper in a coarse fixed
point algorithm to converge on the desired coarse nominal
equilibrium $x_0$; we then proceed as follows (remarkably, the
algorithm is the same for the case of legacy dynamic simulators
and coarse timesteppers of microscopic/stochastic models):

\begin{itemize}

\item   Discretize the domain $D^{n}\subseteq R^{n}$ of the
state-space, where a numerical solution of the feedback
linearization problem is sought in a mesh of, say, N points.
\\

\item Write the transformation vector $S(x)$ as a power series
expansion up to order $p$ around the equilibrium $x_0$ i.e. write
$S(x)$ as $S(x; h)$, where $h \in R^{m}$ is the vector of the
power series coefficients.
For example for a 2-dimensional problem $S(x;h) \equiv
S(x_1,x_2;h)$ can be written as:

\begin{eqnarray}
S_1(x_1,x_2;a_{i=1,...p})=a_1x_1+a_2x_2+\frac{1}{2
!}a_3x_1^2+\frac{1}{2 !}a_4x_2^2+a_5x_1x_2+...+O(p+1)\\
S_2(x_1,x_2;b_{i=1,...p})=b_1x_1+b_2x_2+\frac{1}{2
!}b_3x_1^2+\frac{1}{2 !}b_4x_2^2+b_5x_1x_2+...+O(p+1)
\end{eqnarray}
where:
\begin{equation}
h=[a_1,a_2,...,a_p,b_1,b_2,...,b_p]
\end{equation}

Then, write the feedback control law as in (17).

\item Calculate the values of the unknown coefficients of $S(x;
h)$ using a matrix-free iterative nonlinear solver [Kelley, 1999],
or possibly an unconstrained optimization algorithm, such as the
Broyden, Fletcher, Goldfarb, Shanno (BFGS) method.

\end{itemize}

The optimization problem can be stated as finding the values of
the vector $h$ such that the sum of squared errors on the
discretization mesh is minimized, i.e.:
\begin{equation}
\min_{h} {\frac{1}{2}\sum_{i=1}^{N} \parallel G_{i}(h)
\parallel ^{2} _{2}}
\end{equation}
where the vector function $G_{i}(h)$ is defined as:
$G_{i}(h)=S(\Phi_T(x_{i},-cS(x_{i});h)-AS(x_{i};h)$, $\forall
x_{i}$ on the discretized mesh, and $\parallel \bullet
\parallel _{2}$ is the standard Euclidean norm in the above minimum norm
problem [Luenberger, 1969].

The quantities involved in the optimization computations (e.g. the
values $G_i$) are evaluated repeatedly using the (legacy or
coarse) timestepper for each value $x_i$ in the mesh.
\\
{\bf Remark 6}: The single-step feedback linearization problem
under consideration admits an alternative formulation, where the
inverse transformation map: $x=w(z)$ is sought that satisfies the
following system of NFEs:
\begin{eqnarray}
w(Az)&=&\Phi(w(z), -cz) \nonumber \\
w(0)&=&0
\end{eqnarray}
where:
\begin{equation}
x=w(z)=S^{-1}(z)
\end{equation}
The above functional equation is structurally simpler
(first-order) than (12) (second-order), since in the latter the
unknown vector function $S(x)$ appears through two consecutive
function composition operations. Furthermore, it can be easily
shown that the above problem reformulation leads to the same
results, namely the same feedback linearizing control law
[Kazantzis, 2001]. Notice, that in this case we expand $w(z)$
(instead of $S(x)$) in a power series and we then seek the values
of the vector $h^{'}$ such that the sum of squared errors on the
discretized domain (w.r.t \space z state-space, say
$D^{'n}\subseteq R^{n}$) is minimized, i.e.:
\begin{equation}
\min_{h^{'}} {\frac{1}{2}\sum_{i=1}^{N} \parallel G_{i}^{'}(h^{'})
\parallel ^{2} _{2}}
\end{equation}
where the vector function $G_{i}^{'}(h^{'})$ is defined as:
$G_{i}^{'}(h^{'})=w(Az_{i})-\Phi_T(w(Az_{i}),-cz_i);h^{'})$,
$\forall z_{i}$ on the discretized mesh.

Upon convergence we find the desired transformation $S(x)$
symbolically by applying a functional inverse on $w(z)$.
More generally, matrix-free iterative linear algebra approaches
can be used to solve the discretized nonlinear functional
equations; in these methods the action of the Jacobian is
estimated by appropriately initialized nearby initial conditions
(dictated, for example, by a GMRES protocol).
It is worth noting that, if the problem dynamics are characterized
by a separation of time scales, and the long-term dynamics lie on
a low dimensional, attracting manifold, the dynamical integration
involved in timestepping may be beneficial to the convergence of
such iterative solution techniques [Wington et al., 1985; Kelley
et al., 2004].

\section{An Illustrative Case Study}

\subsection{The Deterministic Version}

Our illustrative example consists of a simplified mechanism for
the dynamics of $NO$ reduction by $H_2$ on $Pt$ and $Rh$ surfaces.
The simplified deterministic mean field model for this mechanism
is given by:

\begin{equation}
\frac{dx}{dt}=\alpha (1-x)-\gamma x-u(1-x)^2 x \equiv L(x,u)
\end{equation}
where $x$ is the coverage of adsorbed $NO$, $\alpha$ is the rate
constant for NO adsorption (incorporating the gas phase $NO$
partial pressure), $\gamma$ is the rate constant for $NO$
desorption, and $u$ is the reaction rate constant (and, in our
scheme, the control variable).
In order to transform the problem back to the discrete time
formulation, we take a forward Euler step of the continuous time
problem
\begin{equation}
x(k+1)=x(k)+TL(x(k),u(k)) \equiv \Phi(x(k),u(k)).
\end{equation}

Simulation results were obtained for: $\alpha = 1, \gamma = 0.01$.
This model, exhibits two regular turning points (at $u\simeq 3.96$
and $u\simeq 26$) as shown in the bifurcation diagram (Figure 2).
We want to derive a nonlinear feedback control based on the
proposed methodology, to stabilize the timestepper at the
open-loop unstable stationary state ($x_0=0.5559,u_0=4$).
We chose $T=0.1$ as the reporting time horizon; the open loop
eigenvalue at the nominal steady state is 0.1459, and the
characteristic time is 6.85; A time step of 0.1 is therefore
sufficient for accuracy of the Euler integration step and
numerical stability.

\subsection{The Microscopic/Stochastic Version}

The procedure remains essentially the same when the timestepper
results are obtained through short bursts of microscopic
simulation.
Here for the stochastic simulations of the mechanism embodied in
(36) we used the Gillespie Stochastic Simulation Algorithm (SSA)
[Gillespie, 1976, 1977].

Given the value of the surface coverage at time $t=0$ we computed
the expected value of the coverage after a reporting time horizon
$T$ by simulating a system with a relatively large number of
available sites (say $N_{size}$), averaging over several
realizations (say $N_{run}$); the system size and number of
realizations were chosen here to be $N_{size}=100^2$ and
$N_{run}=100$, respectively.
The time horizon was again selected to be $T=0.1$.
The Monte Carlo model is considered as a ``black box" coarse
timestepper $x(k+1) = \Phi_{T}(x(k), u(k))$.

The coarse Jacobian (here, a single derivative, which doubles as
the coarse eigenvalue) at the fixed point is estimated by wrapping
a Newton's method around the coarse KMC timestepper.
The coarse identified model (Jacobian and right hand-side) is then
used for tracing the solution branch by coupling to a
pseudo-arc-length continuation scheme [Keller, 1977].
For the continuation we used $N_{size}=200^2$ and $N_{run}=1000$.
For details on the computation of coarse stationary states and
coarse bifurcation diagrams in an equation-free framework see
[Makeev et al., 2002; Gear et al., 2002; Kevrekidis et al. 2003].
The resulting bifurcation diagram coincides with the one obtained
through the deterministic timestepper.
Given the unstable coarse stationary state at $u=4$, the requisite
functional equation for simultaneous feedback linearization and
pole placement was solved using the coarse timestepper and
minimizing Equation (30) using the BFGS method.
To implement this procedure we used deviation variables defined as
$x'=x-x_0$ and $u'=u-u_0$, while $A$ now is a scalar chosen as
0.8.

We derived the unknown transformation map $S(x)$ numerically by
the following distinct ways:

a) Analytically, by expanding $S(x)$ in a power series and
retaining quadratic terms of the form
$S(x)=\alpha_1x+0.5\alpha_2x^2$, substituting $u=-S(x)$ into
$\Phi(x(k),u(k))$ and then expanding $\Phi(x(k),-S(x))$ in Taylor
series around the equilibrium (0,0).
The values of the unknown coefficients $\alpha_i$'s are computed
by equating terms of the same order on both sides of NFE's (12).

b) Equation-free by using the ``black-box" KMC timestepper
approach,  i.e. by solving the optimization problem as appearing
in (31) using the BFGS quasi-Newton method and a line search
technique.

Here the domain of interest was chosen as $D \in R \equiv [-0.1
\hspace{.1in} 0.1]$ and was discretized into 25 equally spaced
points.
In Figure 3 we plot the derived $S(x)$.

The transformation found by solving the NFE using timestepping is
later used to close the loop (simultaneously linearizing and
assigning poles for the closed loop system dynamics)
Figure 4 demonstrates responses resulting from the desired closed
loop dynamics $z(k + 1) =S(k+1)= 0.8 z(k)=0.8 S(k)$ (dotted lines)
and that of the numerically obtained transformation $S(x)$ when
applying the control law  on the coarse KMC timestepper (solid
ones).
Figure 5 shows the {\it closed loop} responses of the
deterministic mean field model and of the Kinetic Monte Carlo
version starting from different initial conditions. These were
obtained through the solution of the norm minimization problem
(31) using the deterministic and the stochastic KMC model
respectively.
The feedback linearizing transformation was found by minimizing
(31) using the BFGS method. The obtained results confirm the
effectiveness of the proposed equation-free nonlinear controller
design methodology, demonstrating successful stabilization and
regulation of the process at the unstable stationary state.

\section{Concluding Remarks}

We demonstrated how feedback linearization with pole placement in
a single step, analyzed in [Kazantzis, 2001] for closed form
equation models, can be performed in an equation-free framework by
acting directly on a fine scale simulator.
The illustrative example used a stochastic realization of a
simplified model of a catalytic surface reaction.
Admittedly, the example is a very simple one; in particular, it is
(coarsely) one-dimensional, and for such systems a feedback
linearization transformation always exists [Isidori, 1999].
Yet the timestepper based, equation-free methodology illustrated
is not restricted to one dimensional (when coarse-grained)
problems; all the elements of the method (the timestepper, the
location of unstable fixed points and their leading slow
eigenvalues, the solution of the corresponding functional
equation) remain effectively the same in higher-dimensional cases.
More sophisticated matrix-free iterative methods can be used to
solve the requisite functional equation by acting directly on the
fine-scale simulator.
Parallel computation (a different replica fine scale simulation of
the same initial condition performed on each processor) and
computational tools like {\it In situ Adaptive Tabulation} (ISAT,
[Pope, 1997]) can be used to alleviate, when appropriate, the
computational wall clock time and effort required to estimate the
necessary coarse-grained quantities.
Furthermore, if a strong separation of time scales (a spectral
gap) appears in the coarse-grained dynamics, and the long-term
behavior lies on a low-dimensional ``slow manifold", it is
possible to take advantage of this through timestepping to solve
an effective NFE of reduced dimension [Gear et al., 2004].
In this paper we assumed that we knew what ``the right"
macroscopic observable was, in which  to restrict the microscopic
system dynamics; the detection of appropriate such observables,
either through data analysis [Coifman et al., 2004] or observer
design is a subject we are currently pursuing.

\newpage
\begin{Large}
{\bf Figure Captions}
\end{Large}

\vspace{0.35in}

{\bf Figure 1}: Schematic of the coarse timestepper in a
controller design framework

\vspace{0.2in}

{\bf Figure 2}: (a) Coarse Bifurcation diagram of the kMC model,
obtained by the coarse timestepper, (b) blow up of the diagram
near the equilibrium of interest; solid lines correspond to stable
coarse steady states while the dotted ones correspond to unstable
coarse steady states

\vspace{0.2in}

{\bf Figure 3}: $S(x)$ as computed analytically (solid line) and
using the black-box coarse KMC timestepper (dotted line)

\vspace{0.2in}

{\bf Figure 4}: Transients of $S(k+1)=0.8 S(k)$, corresponding to
the desired closed loop dynamics, (solid lines) and  $S(x(k))$
using the computed control law  on the coarse KMC timestepper
(dotted lines)

\vspace{0.2in}

{\bf Figure 5}: (a) Transient response for  0.1 initial
perturbation of the coarse state variable from the coarse
equilibrium. (b) Transient response for  0.2 initial perturbation
of the coarse state variable from the coarse equilibrium, (c)
Transient response of the control variable for  0.2 initial
perturbation of the coarse state variable from the coarse
equilibrium (lower ones correspond to -0.2). Simulation runs for
the KMC timestepper were obtained with $N_{size}=100^2$ and
$N_{run}=100$.

\newpage
\begin{titlepage}
\begin{figure}[ht]
\begin{center}
\mbox{\psfig{figure=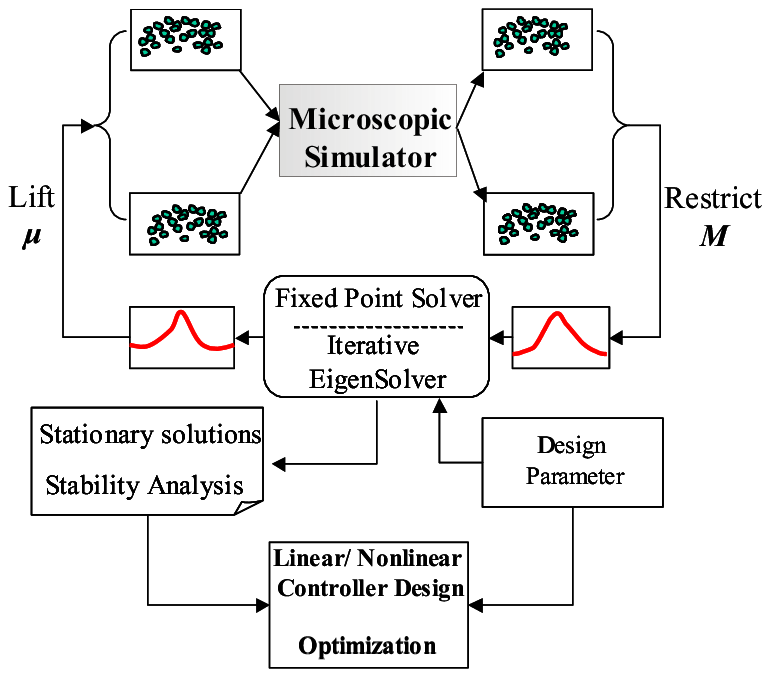, width=19cm}} \caption{Schematic
of the coarse timestepper in a controller design framework}
\end{center}
\end{figure}
\end{titlepage}

\clearpage
\newpage
\begin{titlepage}
\begin{figure}[ht]
\begin{center}
\mbox{\psfig{figure=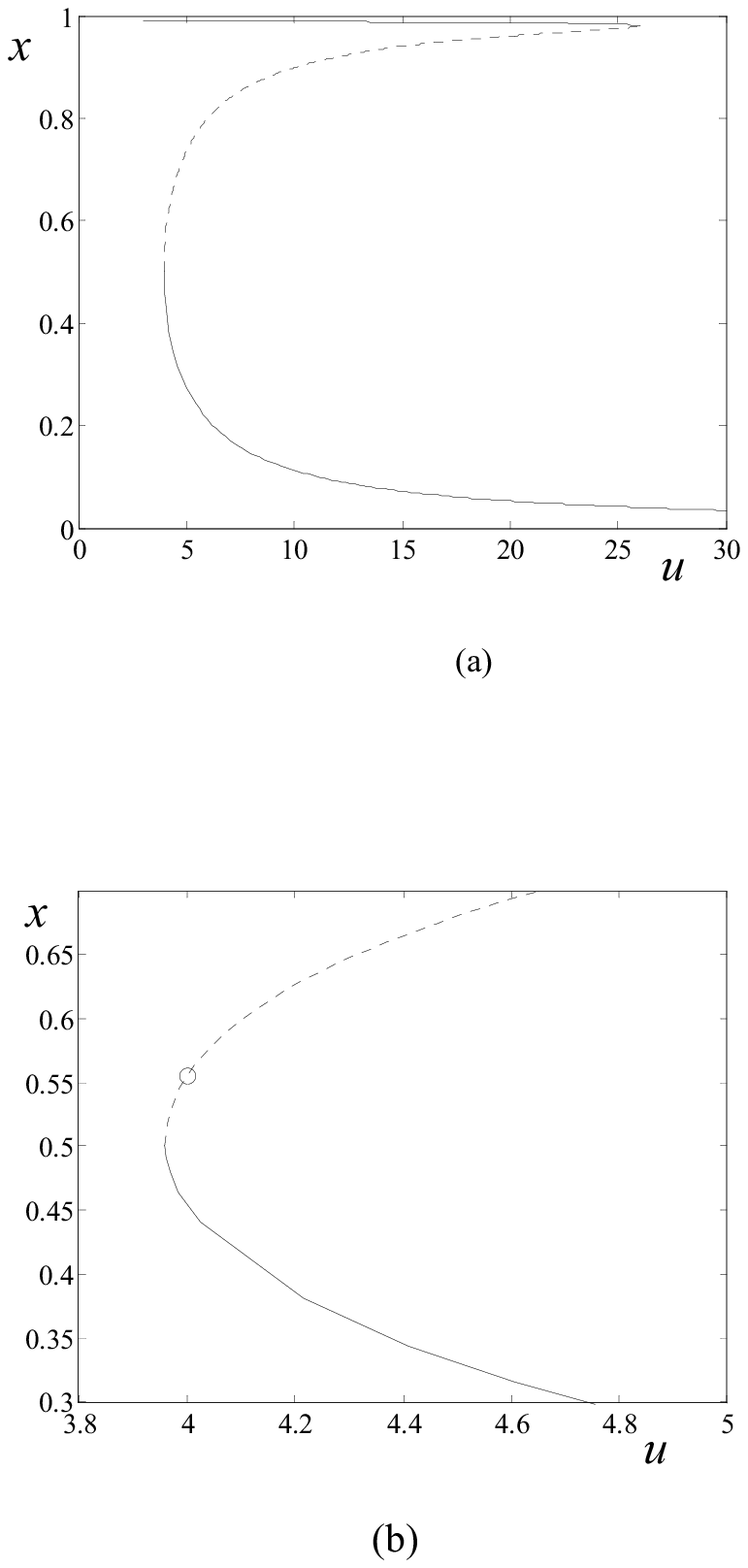,width=18cm}}
\end{center}
\caption{(a) Coarse Bifurcation diagram of the kMC model, obtained
by the coarse timestepper, (b) blow up of the diagram near the
equilibrium of interest; solid lines correspond to stable coarse
steady states while the dotted ones correspond to unstable coarse
steady states}
\end{figure}
\end{titlepage}

\clearpage
\titlepage
\newpage
\begin{figure}[ht]
\begin{center}
\mbox{\psfig{figure=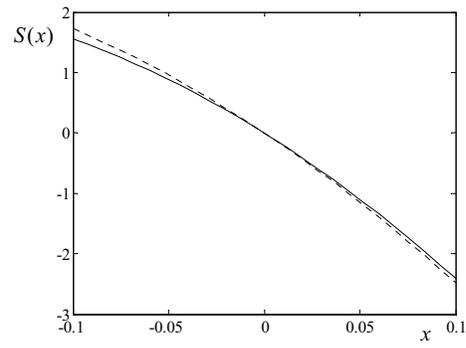,width=18cm}}
\end{center}
\caption{$S(x)$ as computed analytically (solid line) and using
the black-box coarse KMC timestepper (dotted line).}
\end{figure}

\clearpage
\newpage
\titlepage
\begin{figure}[ht]
\begin{center}
\mbox{\psfig{figure=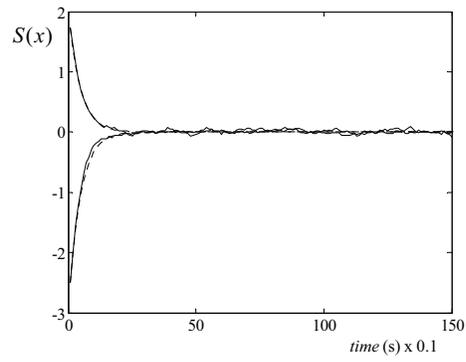,width=18cm}}
\end{center}
\caption{Transients of $S(k+1)=0.8 S(k)$, corresponding to the
desired closed loop dynamics, (solid lines) and  $S(x(k))$ using
the computed control law  on the coarse KMC timestepper (dotted
lines)}
\end{figure}

\clearpage
\newpage
\titlepage
\begin{figure}[ht]
\begin{center}
\mbox{\psfig{figure=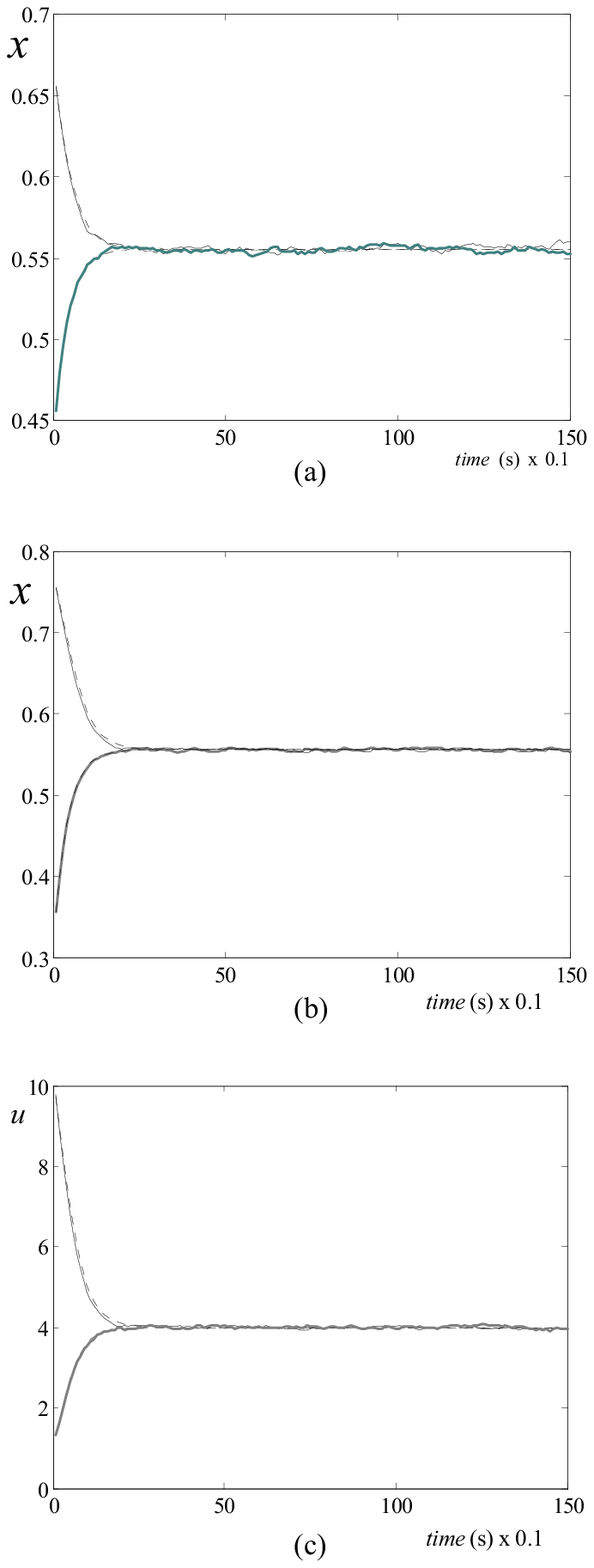,width=18cm}}
\end{center}
\caption{(a) Transient response for  0.1 initial perturbation of
the coarse state variable from the coarse equilibrium. (b)
Transient response for  0.2 initial perturbation of the coarse
state variable from the coarse equilibrium, (c) Transient response
of the control variable for  0.2 initial perturbation of the
coarse state variable from the coarse equilibrium (lower ones
correspond to -0.2). Simulation runs for the KMC timestepper were
obtained with $N_{size}=100^2$ and $N_{run}=100$.}
\end{figure}

\end{large}

\newpage


\begin{thebibliography}{99}
\bibitem{ara} Aranda-Bricaire, E., Kotta, U. and  Moog, C. H.
[1996] ``Linearization of discrete-time systems", {\it SIAM J.
Control Optim.} {\bf 34}, 1999.
\bibitem{arm} Armaou, A., Siettos, C. I. and Kevrekidis, I. G. [2004a] ``Time-steppers and coarse
control of microscopic distributed processes", {\it Int. J. Robust
and Nonlinear Control} {\bf 14}, 89-111.
\bibitem{arm2} Armaou, A., Kevrekidis, I. G. and Theodoropoulos,
C. [2004] ``Equation-free gaptooth-based controller design for
distributed complex / multiscale processes,'' {\it Comp. \& Chem.
Eng.}, {\bf in press}.
\bibitem{cal} Califano, C., Monaco, S. and Normand-Cyrot, D. [1999] ``On the problem of feedback linearization", {\it
Systems Control Lett.} {\bf 36} 61.
\bibitem{car} Carr, J. [1981] {\it ``Applications of Centre Manifold Theory"},
Springer-Verlag, New York.
\bibitem{che} Chen, C. T. [1984] {\it ``Linear System Theory and Design"}, Holt,
Rinehart and Winston, New York.
\bibitem{coif} Coifman, Lafon, R. S., Lee, A. B., Maggioni, M., Nadler, B.,
Warner, F. and Zucker, S.[2004]  ``Geometric diffusions as a tool
for harmonic analysis and structure definition of data", Part I,
Diffusion Maps, {\it Proc.Natl. Acad. Sci.} submitted.
\bibitem{gant} Gantmacher, F. R. [1960] {\it ``The Theory of Matrices"}, Chelsea Publishing Company, New York.
\bibitem{gear} Gear, C. W., Kevrekidis, I. G. and Theodoropoulos, K. [2002] ``Coarse Integration/Bifurcation Analysis via Microscopic
Simulators: micro-Galerkin methods", {\it Comp. Chem. Engng.} {\it
26}, 941-963.
\bibitem{gear} Gear, C. W., Kaper, T. J., Kevrekidis I. G. and Zagaris, A. [2004] ``Projecting on
a Slow Manifold: Singularly Perturbed Systems and Legacy Codes",
{\it submitted to  SIADS, May 2004; can be found as
Physics/0405074 at arXiv.org}
\bibitem{gil1} Gillespie, D. T. [1976] ``A general method for numerically simulating
the stochastic time evolution of coupled chemical reactions", {\it
J. Comput. Phys.} {\bf 22}, 403-434.
\bibitem{gil2} Gillespie, D. T. [1977] ``Exact stochastic simulation of coupled chemical
reactions", {\it J. Phys. Chem.} {\bf 81}, 2340-2361.
\bibitem{gri} Grizzle, J. W. [1986] ``Feedback linearization of discrete-time systems", in: {\it Lecture Notes in Control and Information Sciences}, Springer Verlag, Berlin,
Germany.
\bibitem{guk} Guckenheimer, J. and Holmes, P. [1983] {\it ``Nonlinear Oscillations, Dynamical Systems and Bifurcations of Vector Fields"},
Springer-Verlag, New York.
\bibitem{isi} Isidori, A. [1999] {\it ``Nonlinear Control Systems"}, Springer-Verlag, New York.
\bibitem{jac1} Jakubczyck, B. [1987] ``Feedback linearization of discrete-time systems", {\it
Systems Control Lett.} {\bf 9}, 441.
\bibitem{kaz} Kazantzis, N. [2001] ``A functional equations approach to nonlinear discrete-time feedback stabilization through pole-placement", {\it
Systems Control Lett.} {\bf 43},361.
\bibitem{keller} Keller, H. B. [1977] ``Numerical solution of bifurcation and non-linear eigenvalue problems": {\it in P. H. Rabinowitz (Ed.), Applications of Bifurcation Theory, (Academic Press, New York)}
359-384.
\bibitem{kel} Kelley, C. T., [1999] {\it ``Iterative Methods for Optimization, SIAM series on Frontiers in Applied Mathematics"}, PA.
\bibitem{kel} Kelley, C. T., Kevrekidis, I. G. and Qiao, L., [2004] ``Newton-Krylov Solvers for
Timesteppers", {\it Can be found as Math/0404374 at arXiv.org}
\bibitem{kevr1} Kevrekidis, I.G., Gear, C. W., Hyman, J. M., Kevrekidis, P. G., Runborg, O. and Theodoropoulos,K. [2003] ``Equation-free coarse-grained multiscale computation: enabling microscopic simulators to perform system-level tasks", {\it Comm. Math. Sciences} {\bf 1}, 715-762; original version can be obtained as physics/0209043 at arXiv.org.
\bibitem{kevr2} Kevrekidis, I. G., Gear, C. W. and Hummer, G. [2004] ``Equation-free: the computer-assisted analysis of complex,  multiscale systems", {\it AIChE J.} {\bf 50},
1346-1354.
\bibitem{lee} Lee, H. G., Arapostathis, A. and Marcus, S. I.  [1987] ``On the linearization of discrete-time systems", {\it Int. J. Control} {\bf 45}, 1783.
\bibitem{lin} Lin, W. and  Byrnes, C. I. [1995] ``Remarks on linearization of discrete-time autonomous systems and nonlinear observer design, {\it
Systems Control Lett.} {\bf 25},31.
\bibitem{lue} Luenberger, D. G. [1963] ``Observing the state of a linear system",
{\it IEEE Trans. Milit. Electr.} {\bf 8},74.
\bibitem{luen2}Luenberger, D. [1969] {\it Optimization by Vector
Space Methods}, Wiley, New York.
\bibitem{makeev} Makeev, A., Maroudas, D. and Kevrekidis, I. G. [2002] ``Coarse statbility analysis using stochastic simulators: Kinetic Monte Carlo examples", {\it J. Chem. Phys.} {\bf 116},
10083-10091.
\bibitem{nam} Nam, K. [1989] ``Linearization of discrete-time nonlinear systems and a canonical structure", {\it IEEE Trans. Autom. Contr.} {\bf 34},119.
\bibitem{siet1} Siettos, C. I, Armaou, A., Makeev, A. G., Kevrekidis, I. G. [2003a] ``Microscopic/ stochastic timesteppers and coarse control: a kinetic Monte Carlo example", {it\ AIChE J.} {\bf 49},
1922-1926.
\bibitem{siet2} Siettos, C. I., Graham, M. and Kevrekidis, I. G., [2003b] ``Coarse Brownian Dynamics for Nematic Liquid Crystals: Bifurcation, Projective integration and Control via Stochastic Simulation", {\it J. Chemical Physics} {\bf 118},
10149-10157.
\bibitem{siet3} Siettos, C. I. Maroudas, D. and Kevrekidis, I. G. [2004a] ``Coarse bifurcation diagrams via microscopic simulators: a state-feedback control-based approach", {\it Int. J. Bifurcation and Chaos} {\bf 14}, 207-220.
\bibitem{siet4} Siettos, C. I., Kevrekidis, I. G. and Kazantzis,N. [2004b] ``Nonlinear Feedback Linearization with Pole Placement in One Step:An Equation-free Approach for Discrete Time systems and Microscopic Simulators", {\it Complexity in Science \& Society European Advanced Studies Conference}, Ancient Olympia,
Greece, 22-26 July.
\bibitem{theod} Theodoropoulos, K., Qian, Y. H. and Kevrekidis I.G. [2000] ``Coarse stability and bifurcation analysis using timesteppers: a reaction diffusion example", {\it Proc. Natl. Acad. Sci.} {\bf 97},
9840-9843.
\bibitem{pope} Pope, S. B. [1997]
``Computationally efficient implementation of combustion chemistry
using {\it in situ} adaptive tabulation" {\it Comb. Theor.
Modeling} {\bf 1} 41-63.
\bibitem{boeing} Wington, L. B., Yu, N. J. and Young, D. P. [1985]
``GMRES Acceleration of Computational Fluid Codes", {\it 1985 AIAA
Conference}, 67-74.
\end{thebibliography}
\end{document}